# Direct Comparison of Static and Dynamic Measurements of Spin Generation in a Topological Insulator Thin Film

*Vinay Sharma[1,a)], *Sharadh Jois[1,b)], Ryan Van Haren[1], Gregory M. Stephen[1], M. Tomal Hossain[2],

M. Benjamin Jungfleisch[2], Patrick J. Taylor[3], Aubrey T. Hanbicki[1], Adam L. Friedman[1,c)]

[1] *Laboratory for Physical Sciences, 8050 Greenmead Dr., College Park, Maryland, 20740, U.S.A*

[2] *Department of Physics and Astronomy, University of Delaware, Newark, Delaware, 19716, U.S.A*

[3] *Army Research Laboratory, 2800 Powder Mill Rd., Adelphi, Maryland, 20783, U.S.A*

*These authors contributed equally to this work*

Corresponding Authors: a) vsharma@lps.umd.edu, b) sjois@lps.umd.edu, c) afriedman@lps.umd.edu

## Abstract

The competition between intrinsic spin-orbit physics, magnetic phenomena, and the quality of materials and interfaces governs the charge-to-spin conversion processes that are essential to the implementation of spintronic devices. Direct comparisons of spin parameters, which serve as metrics of device quality, obtained by different measurement techniques are scarce, leading to uncertainty regarding discrepancies and the reliability of the methods. Here, we directly compare the spin Hall coefficient ($\theta_{SH}$) in molecular beam epitaxy grown films of $(Bi_{1-x}Sb_x)_2Te_{3-y}Se_y$ (BSTS, x = 0.58, y = 1) at room temperature using two complementary techniques: a static method using non-local voltage (NLV) measurements in BSTS Hall bars with DC charge current, and a dynamic method using spin-torque ferromagnetic resonance (ST-FMR) measurement in $BSTS/Ni_{80}Fe_{20}$ heterostructures at GHz frequencies. We obtain comparable spin Hall coefficients in angular-dependent ST-FMR ($\theta_{SH} = 4.7 \pm 1.1$) and in NLV ($\theta_{SH} = 2.8 \pm 0.6$). The complex effects of ferromagnetic interfaces while determining spin Hall coefficients using static or dynamic techniques becomes evident by contrasting our results to literature.

**Introduction**

Evaluating new materials for spin-based devices requires understanding spin transport parameters like spin lifetime ($t_s$), spin diffusion length ($\lambda_S$), spin polarization ($P_s$), and spin Hall coefficient ($\theta_{SH}$) [1-3]. These parameters can be derived from various spin transport measurement techniques including ferromagnetic resonance (FMR), FMR driven spin pumping (FMR-SP) [4,5], spin-torque FMR (ST-FMR) [6-8], non-local spin-valves (NLSV) [9,10], harmonic Hall [11], THz generation [12], inverse spin-Hall effect (ISHE) measurements [13], and the spin Seebeck effect [14]. The spin Hall coefficient, a particularly useful parameter, is defined as the ratio of induced spin current and applied charge current ($\theta_{SH} = Js/Jc$) and is a good metric for evaluating new, efficient materials for spintronics. There can be large discrepancies between reports due to differences in measurement technique or sample quality. A comparison of spin transport parameters in topological materials characterized by different techniques is needed to understand discrepancies and find commonalities.

Topological materials are part of a promising class of novel materials for alternate state variable computing that can potentially be used in low-power logic [15,16] and next-generation magnetic random-access memory (MRAM) [17]. In particular, topological insulators (TIs) promise extraordinary access to, and control of, spins through intrinsic spin-momentum locked surface carriers that are topologically protected from backscattering in the semiconducting bulk. Inconsistencies in reported values of spin parameters are found across different measurements of TIs. For the $Bi_{(1-x)}Te_ySb_x$, Khang, et al. reported a giant $\theta_{SH} = 52$ in epitaxial films of $Bi_{0.9}Sb_{0.1}$ using current induced magnetization switching in an adjacent MnGa layer [18]. However, the sputter-grown $Bi_{0.85}Sb_{0.15}$ layer of Ruxian, et al. yielded a lower value of $\theta_{SH} = 6.0$ using second harmonic Hall measurements [19]. A similar discrepancy is seen in $Cd_3As_2$. Stephen, et al. reported $\theta_{SH} = 1.5$ and $\lambda_S = 10$ to 40 µm in epitaxial films of $Cd_3As_2$ using NLSVs and ISHE devices [20]. Wilson, et al. used FMR-SP and ST-FMR measurements in epitaxially grown $Cd_3As_2$ films from a different source and found a lower value of $\theta_{SH} = 0.2$ [21]. It is unclear if these differences are due to the measurement techniques or the sample quality. Quasi-DC transport techniques such as NLSV and ISHE rely on static mechanisms of spin transport. On the other hand, FMR based techniques measure spin dynamics using a ferromagnet and GHz frequency excitations. Direct comparison of static and dynamic techniques is difficult because these studies typically use different materials. This is an impediment to establishing confidence in any particular method or the spin parameters of topological materials.

To integrate spin-based computing with conventional charged-based electronic technologies accurate measurements of spin-charge transduction is critical [22,23]. Most techniques to quantify spin transport properties of a TI require an adjacent ferromagnet (FM) layer to inject or detect spin current. The adjacent FM layer can induce diverse and difficult-to-quantify effects, including interfacial spin transparency, spin-memory loss, and interfacial spin–orbit scattering. Each of these effects may skew the empirical spin parameters solely attributed to the TI. Extending this premise to topological insulators, depositing magnetic materials can lead to new electronic phase transitions [24] that are less compatible with existing spintronic paradigms. Additional precautions need to be taken to minimize such effects by adding tunnel barriers or using techniques that do not require magnetic materials.

In this study, we compare the spin transport parameters measured by both static and dynamic techniques on the same TI sample to control for differences in sample quality, fabrication methods, and experimental approaches. We use the DC non-local voltage (NLV) measurement in Hall bar devices (without FM, thereby eliminating any magnetic interfaces) as the static technique and ST-FMR in coplanar waveguide devices (with FM/tunnel barrier) as the dynamic technique. Both types of devices are fabricated simultaneously on an epitaxial TI film of $(Bi_{1-x}Sb_x)_2Te_{3-y}Se_y$ (BSTS, x = 0.58, y = 1) with standard photolithography techniques, as described in the Methods section. The spin transport is characterized as a function of temperature, magnitude of DC current bias, excitation frequency, and orientation of in-plane magnetic field. Our results show comparable values of $\theta_{SH}$ in both techniques. By contrasting our findings to reported values of $\theta_{SH}$ on different TIs, we notice ST-FMR and harmonic Hall experiments have yielded larger values compared to FMR-SP. Although NLV has been underutilized in the realm of TIs, the agreement of $\theta_{SH}$ between ST-FMR and NLV on BSTS shown here corroborates the accuracy of the measured spintronic parameters despite some differences in their underlying mechanisms. Our results act as a guide for reliable spin transport measurement of future TI samples and indicate removal of sample differences to reduce discrepancies in determination of all spin parameters and increasing confidence in measurements of spin parameters.

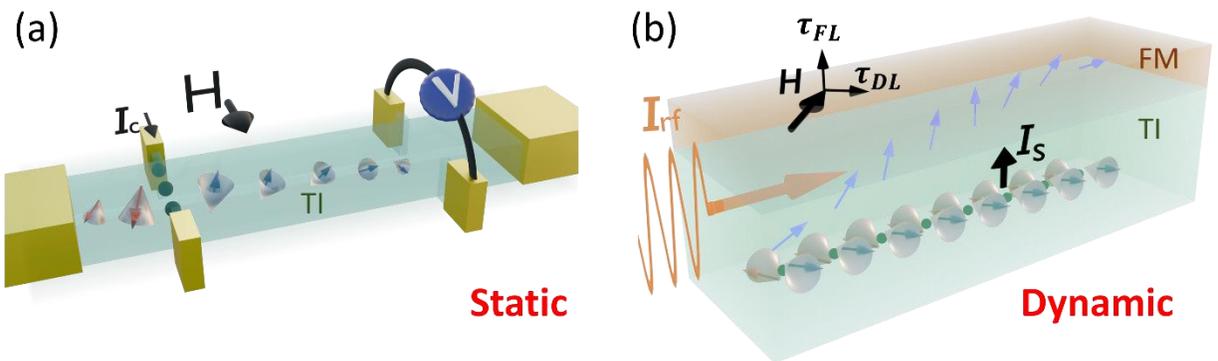

Figure 1 (a) Schematic illustration of static measurement technique (DC non-local voltage) to measure spin-charge interconversion efficiency. Here DC charge current $I_c$ is injected along the Hall contacts which generates a transverse spin current gradient in the TI layer. In presence of magnetic field H, the spin current losses its coherence and starts dephasing, shown by the tilting cones and generate a non-local voltage ($V$), measured at a certain distance from injected Hall leads. (b) Schematic illustration of dynamic measurement technique (ST-FMR) where microwave current ($I_{rf}$) is injected in the TI/FM device at GHz frequency. The charge current flowing through the TI layer generates the spin current $I_s$ which travels towards the FM interface. In experiments, a tunnel barrier (not shown) is required between the TI/FM interface. $I_s$ tunneling into the FM layer generates a spin torque on the magnetic moments of FM layer. Measurements are done at GHz frequencies where the output signal is measured at the resonance field of the FM. A key difference between static DC NLV and dynamic ST-FMR technique is the absence of any FM interface in DC NLV.

**Figure 1** illustrates the measurement techniques. The first case in **Figure 1 (a)**, shows the static technique of injecting a charge current ($I_c$) into the TI layer and measuring a non-local voltage (NLV) a distance away from the charge current path in the TI layer. This method was proposed by Abanin, et al. [25] as a way to avoid FM layers for spin injection or detection and subsequently implemented in experiments [13,26,27]. Films with high spin orbit coupling (SOC) will have a significant transverse spin current via the spin Hall effect (SHE). The spin current is then detected as a non-local voltage at a second set of Hall leads further down the channel due to the ISHE. Sweeping the magnetic field along the direction of current causes high and low resistance states appearing in a pseudo-Lorentzian line shape due to precessional dephasing of the spin current. A minimum in NLV is observed when the applied field matches the resonant field of the system. The schematic in **Figure 1 (b)** illustrates the ST-FMR technique which incorporates an FM interface (with a tunnel barrier, not shown here for simplicity) on the TI. A microwave current ($I_{rf}$) is injected into the device made with a waveguide geometry to reflect the spin signals. The charge current flowing in the TI layer creates a spin current ($I_s$) which flows into the FM layer. It is a dynamic technique where the spin signal is measured in resonance with the applied RF charge current at GHz frequencies. In ST-FMR, the results of interfacial effects at the FM interface can be excluded by angle dependent studies. The angle dependence helps to decouple the in-plane and out-of-plane torques generated at the ferromagnetic resonance frequency and arrive at a reliable spin Hall coefficient.

## I. Methods

Thin films of BSTS were grown to a thickness of $d_{TI} = 12$ nm using 99.9999% pure source materials in a modular molecular beam epitaxy system at a base pressure of $10^{-8}$ Pa. For Se and Te, we use valved cracker sources to obtain a reproducible and highly controlled chalcogenide ratio. For Bi and Sb, we used conventional Knudsen-style cells. Semi-insulating InP (111) wafers were used to seed the (001) orientation of BSTS. The crystalline nature of the films was confirmed using in-situ reflection high-energy electron diffraction (RHEED) and ex-situ high-resolution x-ray diffraction (XRD). Before removal from vacuum, in-situ angle-resolved photoemission spectroscopy (ARPES) measurements capture the electronic dispersion along the Γ − K high symmetry lines of the Brillouin zone (**Figure 2(a)**). The films are capped with a protective layer of Se before removing the samples from the growth chamber to mitigate atmospheric effects from ambient exposure on transport measurements. Additional details on film growth can be found in previous work [28]. This stoichiometry for BSTS (($Bi_{0.42}Sb_{0.58})_2Te_2Se$) was chosen to optimally dope and tune the Fermi level within the bulk band gap. ARPES maps of a pristine BSTS film after growth are shown in **Figure 2(a)**. A clear Dirac cone is visible, and the density of states collapses to a single point at Γ. This is characteristic of a TI. Structural quality and surface roughness are examined using XRD and atomic force microscopy (AFM) measurements and are shown in detail in the supplement (see **supplemental Figure S1 (a) and (b)**). They show a smooth film (surface RMS = $1.32 \pm 0.2$ nm), with a single crystalline nature and nano-crystalline domains.

Hall bar devices for NLV measurements and coplanar waveguide devices for ST-FMR measurements are fabricated simultaneously on the same substrate to treat the BSTS film in all devices identically. See **Figure 2(b) and (c)** for more details about device designs. Prior to device fabrication, the samples are annealed at 200 °C in a high vacuum chamber to remove the Se capping layer. The devices are fabricated using a direct laser write maskless photolithography system (Heidelberg MLA-150). Positive tone exposure resist (AZ-1512) is used for the first exposure to define device channel regions using Ar ion milling. For the following lift-off processes, an additional lift-off resist (LOR 5A) is spun on before AZ-1512. Permalloy ($Ni_{80}Fe_{20}$ or Py, $d_{FM} = 12$ nm) and MgO (5 nm) are deposited by electron beam evaporation only on the ST-FMR devices for spin injection. MgO is used as a tunnel barrier to separate the direct interaction of metallic Py and topological surface states of BSTS. Without MgO, hybridization at the Py/BSTS interface leads to smearing of the topological surface states (See **supplemental Figure S2**). Standard Ti/Au (10 nm/50 nm) contacts are fabricated in the next lithography step to finish all the devices. Thus, the NLV devices have only Ti/Au contacts. Both NLV and temperature dependent ST-FMR experiments were performed in a variable temperature cryostat (Quantum Design PPMS) equipped with a 14 T superconducting magnet. The angle dependent measurements are performed using a high frequency probe station at room temperature.

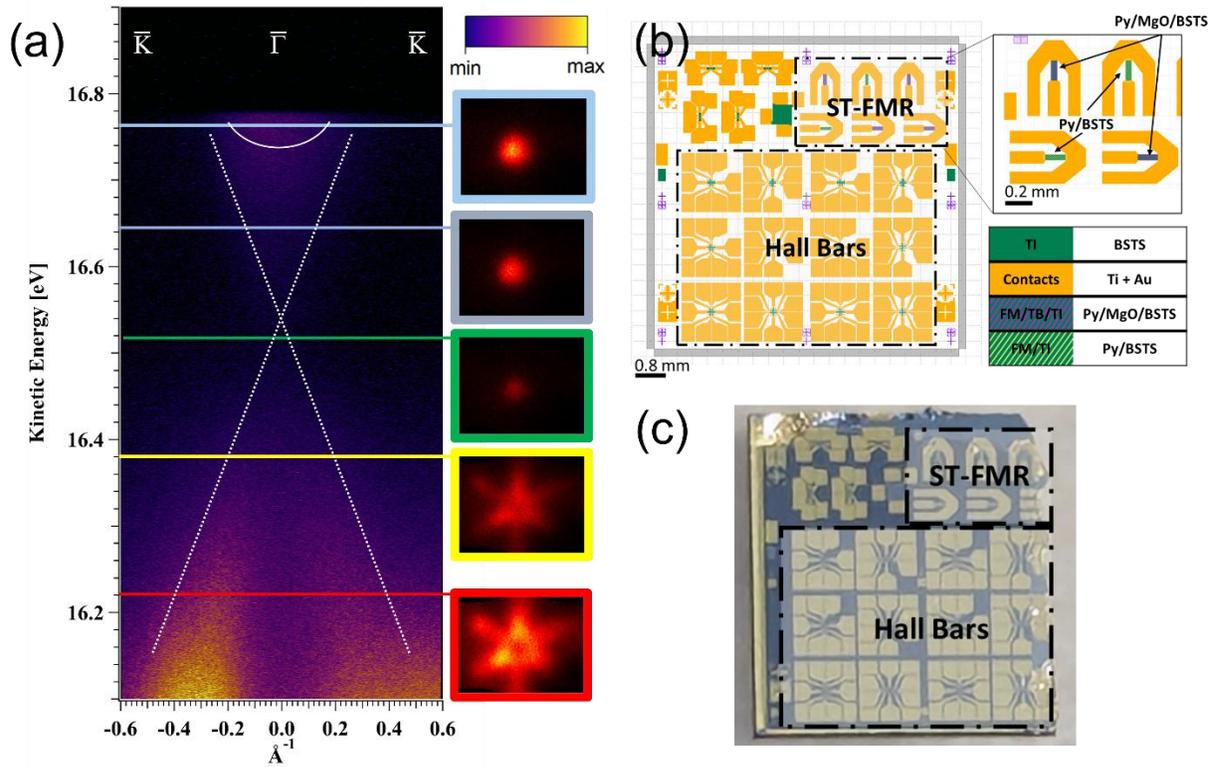

**Figure 2**. (a) ARPES intensity map of BSTS film showing the bulk and surface valence bands along with slices at different energies showing the change or the Fermi surface of the surface states. A faint circular pattern at 16.52 eV (green line) shows the Dirac point. Layout of devices as designed (b) and an image of chip after fabrication (c). In the design layout, the TI (BSTS) layer is in dark green and the metallic contact layers (10 nm Ti, 50 nm Au) appear in gold. The layout is sectioned into two main parts containing co-planar waveguide devices for ST-FMR and Hall bar devices for NLV experiments. The other devices are not relevant for the experiments described here. The ST-FMR section of (b) is magnified to show the two types of devices in orthogonal orientations. Along the diagonals, we see the FM/TI devices made with 12 nm Py and 5 nm MgO tunnel barrier deposited on BSTS. In the off-diagonals, we see the FM/TI devices in which 12 nm Py was deposited directly on BSTS. This design was arrayed to create multiple dies on a single substrate of BSTS/InP to have sufficient identical devices for both NLV and ST-FMR experiments. After fabrication the substrate is diced into several die, such as the one shown in (c).

## II. Results

### A. Thin Film Analysis

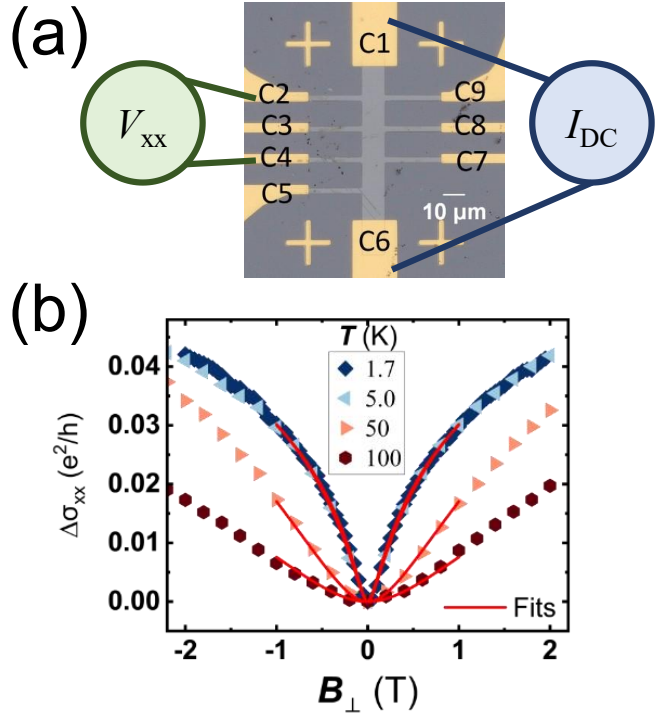

**Figure 3**. (a) Optical image of a Hall bar device with wiring schematics for magnetotransport. (b) HLN fitting of magnetoresistance data (red solid line) with magnetic field applied perpendicular to the plane of the film for different temperatures.

Electrical and magnetotransport is measured using the Hall bar device shown in **Figure 3 (a)**. The temperature dependence of the sheet resistance (**Supplemental Figure S1(c)**) confirms the insulating nature of BSTS in our devices. We observe weak localization (WL) in the magnetotransport at low temperature and fit it to the Hikami-Larkin-Nagaoka (HLN) in **Figure 3(b)**. Additional information on the HLN fit parameters is given in supplemental information and **Supplemental Figure S1(d)**. Typically, weak anti-localization (WAL) is observed in high mobility TI films [28,29] of sufficient thickness. Disorder caused by charge puddles [30], structural non-uniformities [31], or coupling between the top and bottom surface states [32] can result in WL behavior. The nano-crystalline nature of our epitaxial thin film and increased surface roughness seen in XRD and AFM (supplemental **Figure S1(a) and b**)) causes the WL behavior observed here. Due to the disorder in the film, we observe low mobility of $\mu = 50$ cm$^2$/V s at room temperature. Nonetheless, the observation of localization behavior and ARPES measurements confirms 2D transport behavior in our TI films.

The BSTS films used in this study are highly resistive at low temperature. Thus, low temperature (T < 100 K) ST-FMR and NLV measurements are unreliable (**supplemental Figure S3**). Therefore, the spin transport measurements shown in the main text are conducted near room temperature where bulk conduction and SOC is present.

### B. NLV Devices and Measurements

As discussed earlier, for static transport characterization without a FM contact we use the device shown in the schematic of **Figure 4(a)** When a charge current is injected between a pair of contacts on the side of a Hall bar, a transverse spin current is spontaneously generated by the SHE. The spin-up (blue arrow) carriers and spin-down (red arrow) carriers diffuse in opposite directions. The propagation of pure spin current accumulates charge due to the ISHE and is measured as a non-local voltage completely outside the charge current path. Applying a sweeping external in-plane magnetic field $(B_y)$ results in a pseudo-Lorentzian dependence of the non-local resistance $R_{nl}$ due to precessional dephasing (Hanle effect). We fit the non-local resistance using Eq. **1** below to determine the spin diffusion length $\lambda_s$, spin relaxation time $t_s$, and the spin-Hall coefficient $\theta_{SH}$:

$$\Delta R_{nl}(B_y) = \frac{A}{\lambda_s} Re\left[\sqrt{1 + i\,HG} \exp\left(-\frac{L}{\lambda_s}\sqrt{1 + i\,HG}\right)\right] + c_1(H)^2 + c_2(H) \qquad \textbf{Eq. 1}$$

The dominant fitting parameter is the spin diffusion length $\lambda_s$. The spin relaxation time is captured in $G = \Gamma t_s$, where, $\Gamma = \frac{ge}{2m^*}$ is the gyromagnetic ratio, $g = 2$ is the Landé g-factor, and $m^* = 0.30 \pm 0.05\ m_e$ is the effective mass used here based on the cyclotron mass reported in literature. The uncertainty in effective mass contributes to the uncertainty in fits of $t_s$. The pre-factor $A = \frac{1}{2}\theta_{SH}^2 \rho_{sheet} W$, where $\rho_{sheet}$ is the temperature dependent sheet resistance, $W$ is the channel width and $L$ is the separation between the non-local contacts. The spin-Hall angle $\theta_{SH} = \frac{J_s}{J_c}$ is a ratio of the generated spin current ($J_s$) to the supplied charge current ($J_c$). The parabolic and linear terms with pre-factors $c_1$ and $c_2$, respectively, are also added in **Eq. 1** to fit the background magnetoresistance and thermal contributions that become prominent with increasing bias, as we will discuss below. Advantageously, this method does not require the use of tunnel barriers and magnetic contacts common in other methods, greatly simplifying fabrication, measurement, and analysis [20,25,26]. However, it does not allow us to distinguish between surface and bulk effects. Also, care must be taken to eliminate possible spurious Hall signals and geometric resistance. Additional information is available in the literature [13,27].

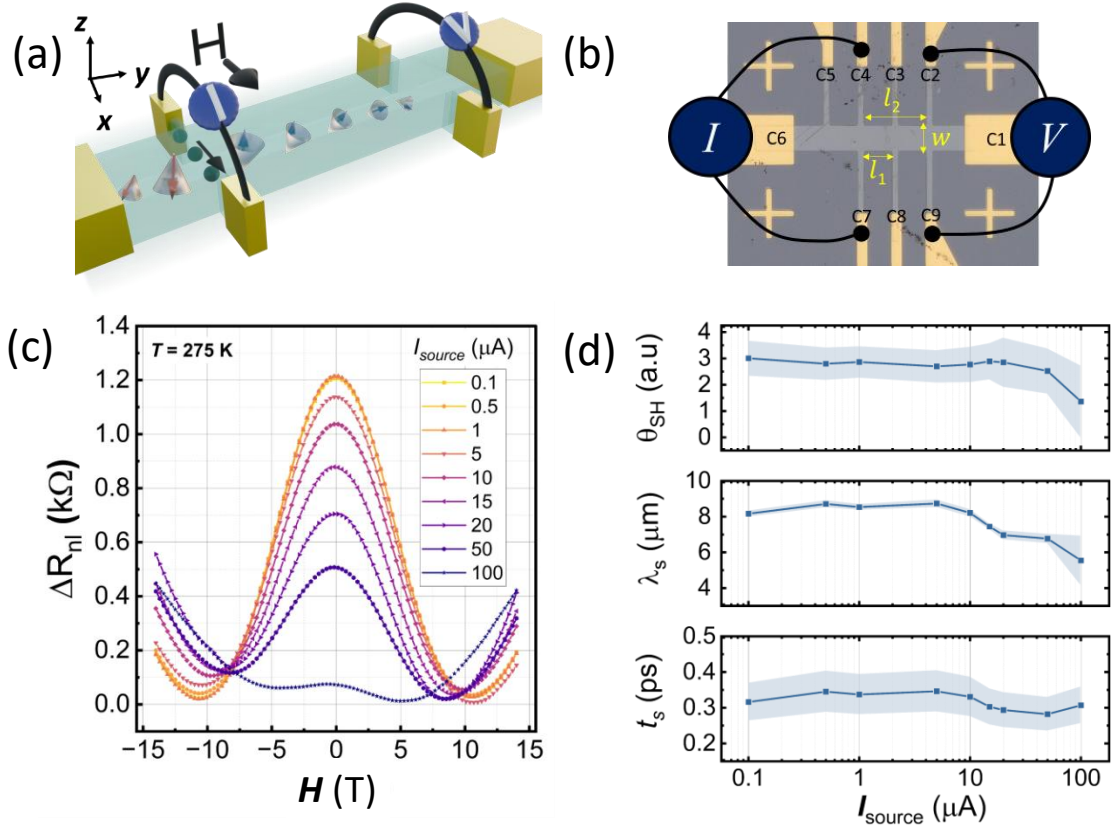

**Figure 4**. Schematic representation of non-local voltage measurement of spin transport. (b) Hall bar device on BSTS film with wiring schematics used to measure the non-local voltage. The dimensions here are $l_1$=14 μm, $l_2$=28 μm and $w$=10 μm. (c) The non-local resistance as a function of in-plane magnetic field ($B_x$) at different current bias. (d) The fit parameters from Eq.1 yield the spin Hall efficiency ($\theta_{SH}$), spin diffusion length ($\lambda_s$), and spin relaxation time ($t_s$) as a function of bias current.

An example of a Hall bar device on BSTS with several contacts and the circuit for the NLV measurement is shown in **Figure 4(b)**. Current is injected between contacts C4 and C7 and two non-local voltages are measured along the length of the Hall bar. The first NLV between C3 and C8 ($l_1 = 14$ μm, w = 10 μm) is too close to the injector contacts $\left(\frac{l_1}{w} = 1.4\right)$, and showed leakage of parabolic magnetoresistance background into the Hanle signals. These data are in the **supplemental Figure S4**. Abanin et al. [25] suggested $\frac{L}{W} \geq 3$ to attain a pure spin signal. The NLV measured between C2 and C9 ($l_2 = 28$ μm, $\frac{l_2}{w} = 2.8$ ) is shown in **Figure 4(c)** as a function of field applied in the direction of the charge current for different current biases at $T = 275$ K. We observe the pseudo-Lorentzian $R_{nl}$ response to sweeping in-plane field at all currents. The maximum signal appears when $I_{source} < 1$ μA and monotonically decreases as the bias current increases. The NLV signal is nearly extinguished at $I_{source} = 100$ μA. The background linear slope and parabolic response also increases as $I_{source} > 5$ μA. The large currents require a high voltage at the current injecting contacts resulting in a large electric field resulting in band bending at the contacts. Together these effects lead to suppression of the spin channel due to competition from increased charge current.

The spin parameters $\theta_{SH}$, $\lambda_s$, and $t_s$ from fitting the experimental data are shown, respectively, in the top, middle, and bottom panels of **Figure 4(d)**. The points are values derived from the fits, and the shaded areas represent the errors in the fitting. The spin Hall coefficient is $\theta_{SH} = 2.8 \pm 0.6$ when $I_{source} \leq$ 15 µA and decreases to $\theta_{SH} = 2.5 \pm 0.9$ at $I_{source} = 50$ µA. The spin diffusion length $\lambda_s = 8.5 \pm 0.2$ µm up to $I_{source} = 10$ µA, then above 15 µA, starts to decrease. The spin relaxation time remains at $t_s = 0.30 \pm 0.05$ ps as bias is increased. The error band for $\tau_s$ is due to the uncertainty in the free electron mass, $m^* = 0.30 \pm 0.05\ m_e$. The values for spin-Hall coefficient observed here are consistent with previous reports on static experiments [20,27]. There is a large error in determining $\theta_{SH}$ at 100 µA because of the low NLV seen in **Figure 4(c)**, attributed to high thermal scattering that reduces charge-to-spin conversion, similar to previous work [9,20,27].

## C. ST-FMR Devices and Measurements

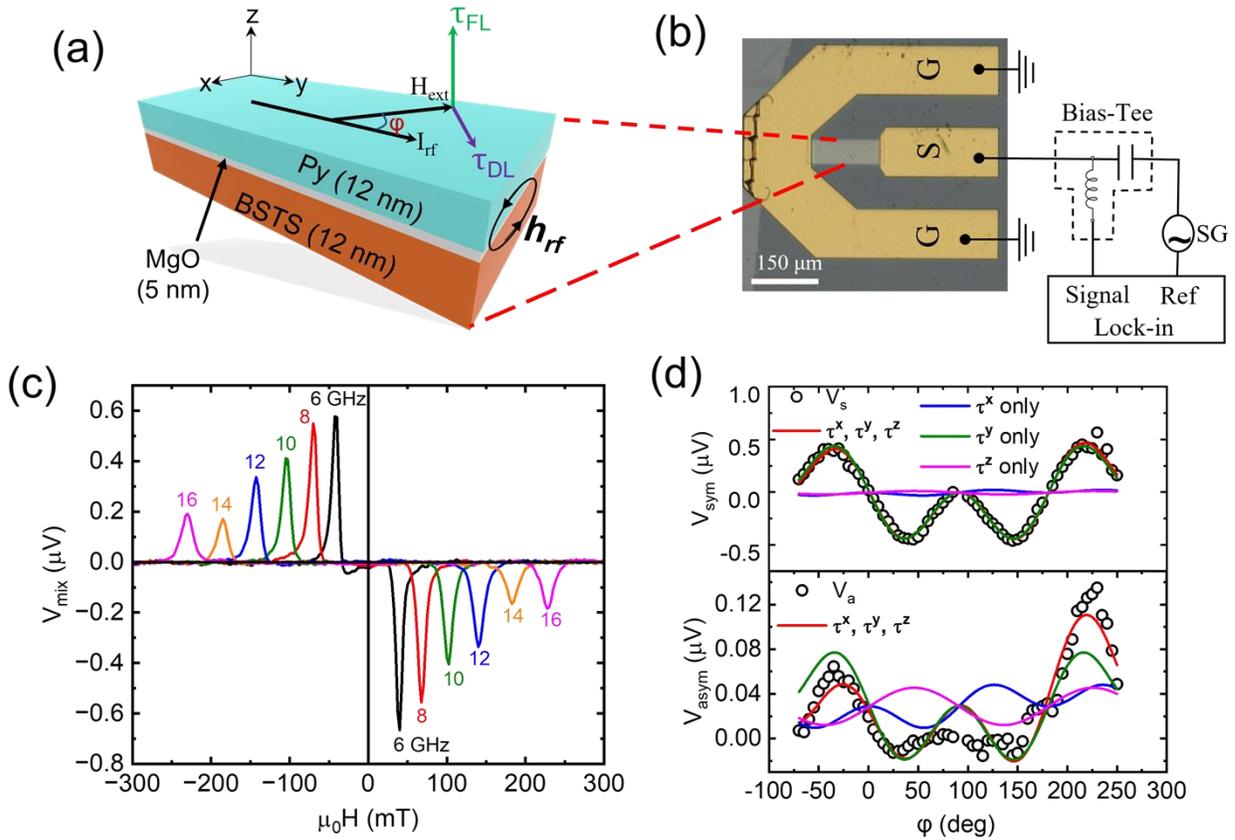

**Figure 5.** (a) Schematic cross-section of the BSTS-MgO-Py (12 nm/5 nm/12 nm) ST-FMR device and (b) circuit used for the ST-FMR experiment. The "G" and "S" annotations on the device stand for ground and signal respectively. The coordinate axes are given to show the direction of in-plane ($\tau_{//}$) and out of plane ($\tau\perp$) torques. An in-plane external magnetic field ($H_{ext}$) is applied at angle of φ with respect to the direction of RF current ($I_{rf}$). (c) The ST-FMR signal measured at φ=45° using driving frequencies from 6-16 GHz at T = 285 K. (d) The angular dependence taken using an excitation frequency of 10 GHz at room temperature shows the contribution of spin torques in all directions to extract the spin Hall coefficient using Eq. 3. The symmetric (upper panel) and anti-symmetric (lower panel) rectifying voltages are shown as a function of φ. The solid blue, green and magenta lines are the fit curves corresponding to the torques acting in x, y and z directions. The red curve is the sum of all torques.

ST-FMR is a dynamic method for extracting spin parameters including the spin Hall angle, $\theta_{SH}$. As shown in **Figure 5(a)**, we apply an in-plane magnetic field ($H_{ext}$) at different angles ($\phi$) with respect to the direction of supplied radio frequency (RF) current ($I_{rf}$) at different excitation frequencies ($f$). **Figure 5(b)** shows a micrograph of the co-planar wave guide device used in the ST-FMR experiments. An in-plane magnetic field is applied at an angle $\varphi$ with respect to the direction of in-plane RF current. The RF current in BSTS generates a spin current in the z-direction, which tunnels through the MgO barrier and excites spins precession in the Py layer. The intrinsic anisotropic magnetoresistance (AMR) of Py oscillates at the same RF frequency as $I_{rf}$ and produces a DC voltage $V_{mix}$, measured using a lock-in amplifier. **Figure 5(c)** shows the representative $V_{mix}$ signals for the BSTS/MgO/Py device taken at 285 K for positive and negative magnetic fields, corresponding to $\varphi = 45°$ and $\varphi + 180°$, collected at different driving frequencies ranging from $f = 6$ to 16 GHz. The measured $V_{mix}$ is a superposition of symmetric ($V_{sym}$) and antisymmetric ($V_{asym}$) components of the ST-FMR signal. The values of $V_{sym}$ and $V_{asym}$ are determined by fitting $V_{mix}$ signal to **Eq. 2** below [33],

$$V_{mix} = V_{sym}\frac{\Delta H^2}{\Delta H^2 + (H_{ext} - H_r)^2} + V_{asym}\frac{\Delta H(H_{ext} - H_r)}{\Delta H^2 + (H_{ext} - H_r)^2} \quad \text{Eq. 2}$$

Here, the FMR linewidth ($\Delta H$), resonance field ($H_r$), and respective amplitudes $V_{sym}$ and $V_{asym}$ are treated as fitting parameters and extracted at each frequency (**supplemental Figure S5**). We determined the effective magnetization ($\mu_0 M_{eff}$) and Gilbert damping parameter (α) by fitting frequency dependence of $H_r$ and $\Delta H$, respectively (see **supplemental Figure S6**). In ST-FMR, the symmetric component $V_{sym}$ is proportional to the damping-like (DL) torque ($\tau_{DL}$) and the anti-symmetric component $V_{asym}$ is proportional to the field-like (FL) torque ($\tau_{FL}$). The ratio of these two torques give the value of charge-to-spin conversion efficiency, or spin-Hall coefficient [6]

$$\theta_{SH} = \frac{V_{sym}}{V_{asym}}\left(\frac{e\,\mu_0\,M_s\,d_{FM}\,d_{TI}}{\hbar}\right)\left[1 + \left(\frac{\mu_0 M_{eff}}{H_r}\right)\right]^{1/2} \quad \text{Eq. 3}$$

where $d_{FM}$ and $d_{TI}$ are the thickness of Py and BSTS, respectively. Based on the line-shape analysis method described above, we can estimate the value of $\theta_{SH}$ irrespective of $I_{rf}$ or the sample resistance. Using **Eq. 3** above, we find $\theta_{SH}$ ranges between $6.1 \pm 0.5$ to $11.5 \pm 0.5$ depending on the variation of RF power with the change in excitation frequency. The lower value corresponds to the maximum RF power and lowest frequency we used, and the higher value corresponds to the lowest RF power and highest frequency. There is a monotonic increase in $\theta_{SH}$ with increase in frequency which signifies that the role of magnetization precession angle varies with excitation frequency.

Overestimation of the spin-Hall angle is a major concern in these measurements. Oersted field-like torque is not completely dependent upon the out-of-plane torque components. It can vary with the sample thickness, magnetization, or unexpected dead magnetic areas at the interface. To further substantiate the frequency dependent results of $\theta_{SH}$, we studied the angular dependence of the ST-FMR signal. **Figure 5(d)** shows the angle dependent ST-FMR data with microwave current and $H_{ext}$ applied in-plane to the sample at different angles of $\varphi$ from $-90°$ to $270°$. Here, $\tau^x$, $\tau^y$, and $\tau^z$ correspond to the torques acting along the respective coordinate axes shown in the schematic of **Figure 5(a)**. These torques show different angular dependencies that are proportional to $V_{sym}$ and $V_{asym}$ arising due to the different torques on the magnetization. The generalized form of symmetric and antisymmetric components is given by [34]

$$V_{sym} \propto \sin(2\varphi)[\tau^x_{DL} \sin\varphi + \tau^y_{DL} \cos\varphi + \tau^z_{FL}] \quad \textbf{Eq. 4}$$

$$V_{asym} \propto \sin(2\varphi)[\tau^x_{FL} \sin\varphi + \tau^y_{FL} \cos\varphi + \tau^z_{DL}] \quad \textbf{Eq. 5}$$

Again, DL refers to the damping-like and FL refers to the field-like torques. As expected by convention, the fits for $V_{sym}$ and $V_{asym}$ from **Eq. 4** and **Eq. 5** shown in Figure 5(d) are consistent with $\tau_y$ (green lines), as well as with all the torques combined (red). Furthermore, we calculate the anisotropic charge-to-spin Hall coefficients by taking the ratio of $\tau^x_{DL}$, $\tau^y_{DL}$, and $\tau^z_{DL}$ with $\tau^y_{FL}$ using

$$\theta^x_{zx} = \frac{\tau^x_{DL}}{\tau^y_{FL}} \frac{e\mu_0 M_s td}{\hbar} \left[1 + \left(\frac{\mu_0 M_{eff}}{H_r}\right)\right]^{1/2} \quad \textbf{Eq. 6}$$

$$\theta^y_{zx} = \frac{\tau^y_{DL}}{\tau^y_{FL}} \frac{e\mu_0 M_s td}{\hbar} \left[1 + \left(\frac{\mu_0 M_{eff}}{H_r}\right)\right]^{1/2} \quad \textbf{Eq. 7}$$

$$\theta^z_{zx} = \frac{\tau^z_{DL}}{\tau^y_{FL}} \frac{e\mu_0 M_s td}{\hbar} \quad \textbf{Eq. 8}$$

Here, the superscripts x, y, and z represent the direction of spin orientation and subscripts zx represents the plane of spin motion. The anisotropic spin Hall coefficients are $\theta^x_{zx} = 0.17 \pm 0.02$, $\theta^y_{zx} =$ $4.71 \pm 1.06$, and $\theta_{zx}^{z} = -0.04 \pm 0.005$ for spin currents with spin polarizations and, hence, torques in $x$, $y$, and $z$ directions, respectively. The results clearly show that charge to spin conversion is predominantly in the $y$-polarized spin channel, as one would expect from conventional spin-Hall effect. The resulting $\theta_{zx}^{y}$ component correctly represents the $\theta_{SH}$ from ST-FMR after accounting for anisotropic torques that occur at ferromagnetic interfaces.

## III. Discussion

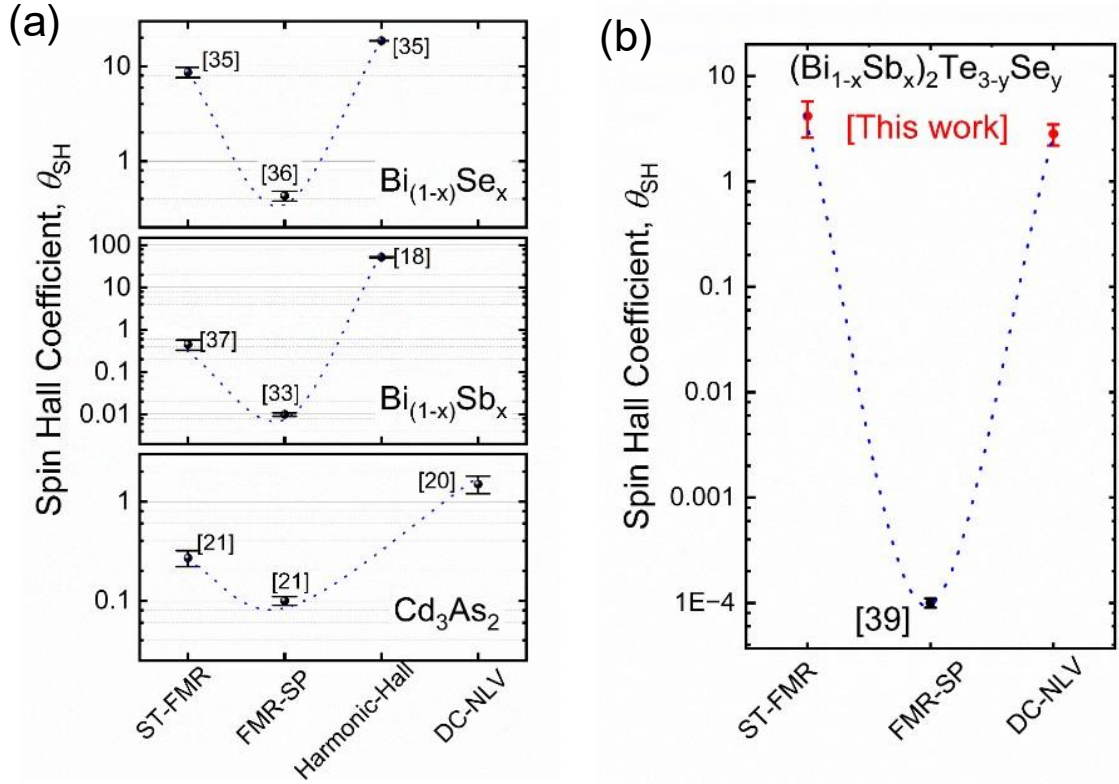

Figure 6. (a) A comparison of spin Hall coefficient measured by different techniques in various materials having high SOC. (b) Variation of spin Hall coefficient in BSTS film using DC NLV and ST-FMR compared with literature measurements using FMR-SP. Numbers inside square brackets are the references. The dotted lines are a guide to the eye.

We compare the results obtained above to reports on thin films of different topological materials at near room temperature based on static and dynamic techniques. **Figure 6(a)**, shows $\theta_{SH}$ for $Bi_{1-x}Se_x$, $Bi_{1-x}Sb_x$, and $Cd_3As_2$, in three panels based on two dynamic techniques (ST-FMR and FMR-SP) and two static techniques (harmonic Hall and DC-NLV). In the top panel, Mahendra DC. et al. [35] found a giant spin Hall coefficient in $Bi_{1-x}Se_x$ using ST-FMR ($\theta_{SH} = 9 \pm 1.08$) and harmonic Hall measurements ($\theta_{SH} = 18 \pm 0.13$) and attributed it to the quantum confinement and induced Rashba-like effect in nanocrystalline

BiSe films. Later, the same research group found lower spin Hall coefficient ($\theta_{SH} = 0.43 \pm 0.05$) in the similar composition of BiSe films [36] using FMR-SP. For $Bi_{1-x}Sb_x$ films in the middle panel, Khang et al. reported a gigantic spin Hall coefficient ($\theta_{SH} \sim 51$) measured using harmonic Hall measurements [18]. Contrarily, lower values were seen in $Bi_{1-x}Sb_x$ by ST-FMR ($\theta_{SH} \sim 0.45 \pm 0.12$) by K. Kondou et al., [37] and by FMR-SP ($\theta_{SH} \sim 0.01 \pm 0.001$) [33]. In the bottom panel, the spin hall coefficient ($\theta_{SH} = 1.5 \pm 0.3$) measured in $Cd_3As_2$ using DC-NLV measurements [20] is an order of magnitude higher than ST-FMR or FMR-SP measurements [21]. This difference can be attributed to the sample quality and imperfect interface of $CdO_x$ with Py used in the FMR experiments. Despite differences in samples, lower values of $\theta_{SH}$ are encountered in dynamic techniques of measuring spin transport in topological materials most likely primarily due to the quality of magnetic interfaces. This is further supported by our findings on our control devices of Py/BSTS (**supplemental Figure S2**) mentioned earlier. The extremely large values of $\theta_{SH}$ seen in the static harmonic Hall experiments on $Bi_{1-x}Se_x$ and $Bi_{1-x}Sb_x$ are possibly due to the intrinsic anomalous Hall signal of the ferromagnet layer [38]. However, more modest values are seen in the NLV experiments on $Cd_3As_2$ without ferromagnet contacts that are corroborated with spin valve experiments in the same work [20]. This comparison further supports the complex role of magnetic interfaces while determining $\theta_{SH}$ in TIs.

Our detailed examination between NLV (static, no FM) and ST-FMR (dynamic, with FM tunnel barrier) yielded comparable values of $\theta_{SH}$ in BSTS, shown in **Figure 6(b)**. In the static experiments on BSTS, we focus on the voltage measured farthest away from the contacts sourcing charge current, i.e., $l_2 = 28$ μm to ignore background contributions from magnetoresistance and competition from charge carriers. The results here are $\theta_{SH} = 2.8 \pm 0.6$ at 275 K. Our previous low temperature study on BSTS [27] found $\theta_{SH} \sim 1$. The ST-FMR measurements yielded a range of values with a high $\theta_{SH} \sim 11.5$, supported by large symmetric Lorentzian components ($V_{sym}$) in the $V_{mix}$ signal taken at different frequencies. This is overestimated because of the strong relationship between of out-of-plane torque and Oersted field generation. To remove those effects, we used angle dependent ST-FMR measurements and found $\theta_{SH} = 4.7 \pm 1.1$ at 10 GHz, likely an upper bound for the spin Hall coefficient and comparable to our static result.

The large $\theta_{SH}$ was not observed when the Py layer was in direct contact with BSTS due to FM coupling smearing the TSS. This aspect is further emphasized by noticing the small $\theta_{SH} = 10^{-4}$ in a previous report using FMR-SP [39], shown in **Figure 6(b)**. The MgO tunnel barrier used in our ST-FMR devices acts as a spin filter [40] and shows the importance of decoupling FM and TI interfaces to obtain reliably high $\theta_{SH}$ measurements.

The two different techniques used here to measure the same BSTS sample provide comparably large $\theta_{SH}$, consistent with expectations of TIs based on previous reports. The nanocrystalline nature of the BSTS films can enhance the quantum confinement effect and lead to the higher $\theta_{SH}$ seen here, similar to granular films of $Bi_xSe_{1-x}$ [35].

## IV. Conclusion

In summary, we measured the room temperature spin Hall coefficient of an optimally doped topological insulator thin film using two complimentary techniques: static DC NLV and dynamic ST-FMR. In DC NLV, we used a Hall bar geometry to inject charge current and found $\theta_{SH} = 2.8 \pm 0.6$, with variation correlated to injected charge current. In ST-FMR, we measured $\theta_{SH} = 4.7 \pm 1.1$ using a non-interacting ferromagnetic interface enabled by a thin MgO tunnel barrier between the FM and TI. Spurious effects are eliminated by using angular measurements and a careful analysis of all the in-plane and out-of-plane torques. Lastly, we compared the spin Hall coefficient in different TIs based technique reported in the literature, emphasizing the detrimental effects of direct magnetic interfaces on TIs in dynamic techniques. Our results provide a guideline to confidently measure the spin Hall coefficients using static or dynamic techniques in topological materials, crucial for developing novel spintronic devices.

## V. Acknowledgments

Work at the University of Delaware including angular-dependent ST-FMR measurements was supported by NSF through the University of Delaware Materials Research Science and Engineering Center, DMR-2011824.

# Supplemental Information

## I. Magnetotransport and Structural Quality

The structural quality of the BSTS film is established by X-ray diffraction (XRD) measurements. Figure 3 **(a)** shows the results of grazing angle incidence XRD, showing clear 000n peaks of BSTS, suggesting the c-axis oriented growth. Three of the BSTS peaks lead to a lattice constant of c ~ 29.7 Å, close to the hexagonal phase of BSTS [1]. The broad peaks signify a fairly small nano-crystallite size approximately 6.30±0.5 nm, suggesting a nano-crystalline nature of the film. No other phase of polycrystalline BSTS were observed in XRD maps. **Figure S1(b)** shows the surface topography of BSTS film with surface roughness = 1.32±0.2 nm. The nanocrystalline nature of BSTS films is visible in this topography.

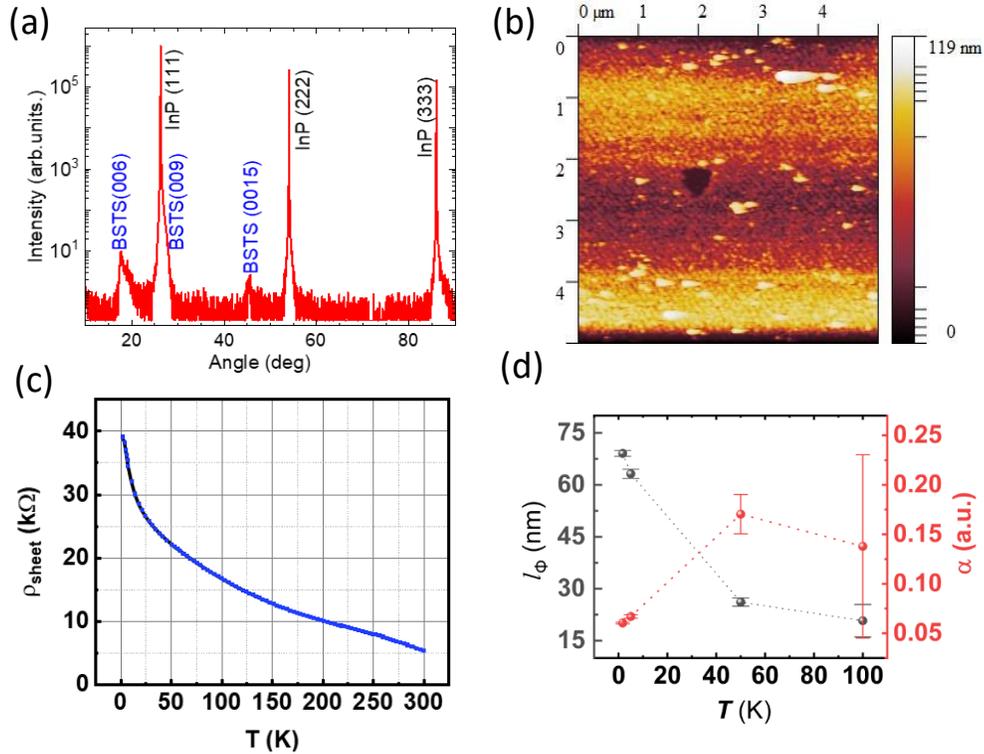

**Figure S1**. (a) X-ray diffraction from a BSTS film in grazing angle incidence. All three BSTS peaks arise from 000n family of hexagonal structure. (b) Surface topography of BSTS film imaged using AFM. Surface roughness = 1.32±0.2 nm. (c) The phase coherence length ($l_\varphi$) and α extracted after HLN fitting at different temperatures are shown here. (d) The sheet resistance as a function of temperature for the thin film.

Below we provide the basic properties of the BSTS thin film using the connections shown in the main text Figure 3 (a). The sheet resistance of the film shows a gradual increase in resistance as temperature is lowered, in **Figure S1 (c)**. We fit the WL in the differential longitudinal conductivity using the Hikami-Larkin-Nagaoka equations given by [2]

$$\frac{\Delta\sigma_{xx,\perp}}{\sigma_0} = \frac{\alpha e^2}{2\pi h}\left[\psi\left(\frac{H_\perp}{H_z}+\frac{1}{2}\right) - \ln\left(\frac{H_\perp}{H_z}\right)\right] \text{ and } H_\perp = \frac{\hbar}{2el_\phi^2} \quad \textbf{\textit{Eq. S1}}$$

where $\psi$ is the Digamma function, $l_\phi$ is the phase coherence length, and e, h, and $\hbar$ are the electron charge, Planck constant, and reduced Planck constant, respectively. The prefactor, $\alpha$, is a constant and is expected to be ~0.5 for each Dirac surface state in bulk topological insulators (TI). Our results in Figure 3 **(d)** show $\alpha$ ranges between 0.05 to 0.15. The small $\alpha$ observed here is attributed to thin films where there can be coupling of surface states.

## S2.  ST-FMR in BSTS/Py sample at room temperature

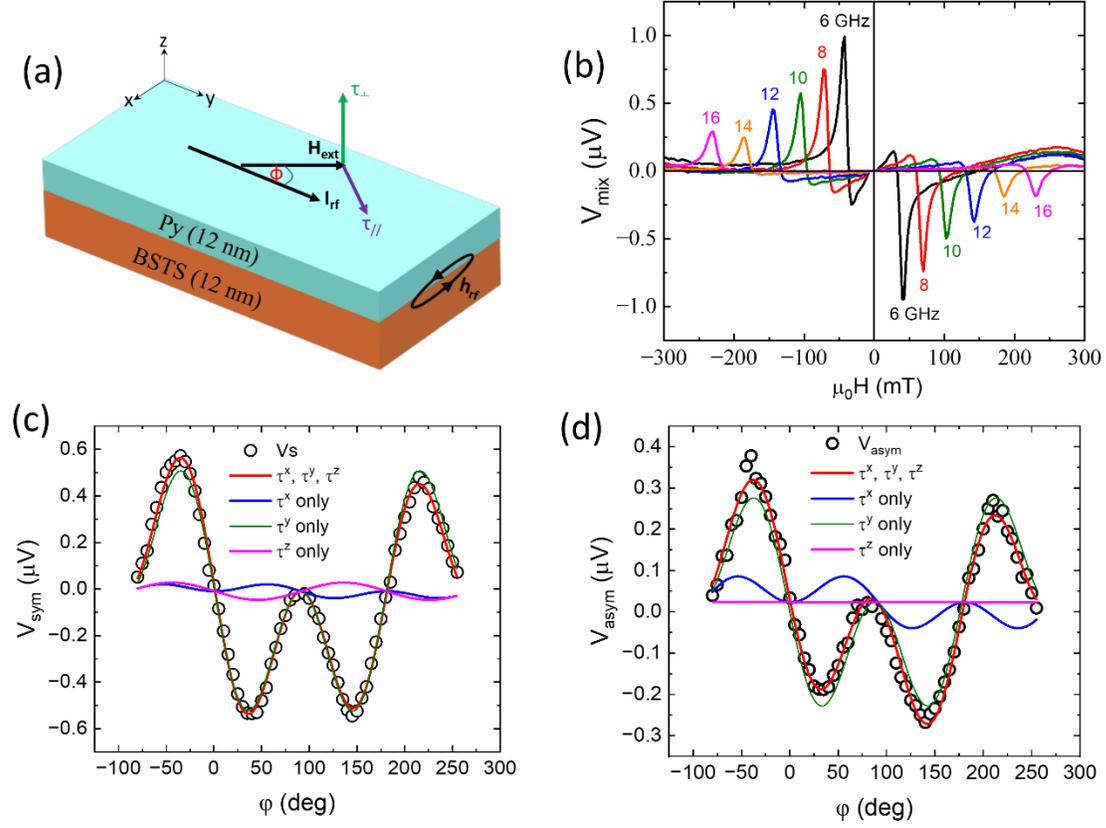

**Figure S2.** (a) Schematic cross-section of the BSTS-Py (12 nm/12 nm) device used for the ST-FMR experiment. The coordinate axes are given to show the direction of in-plane ($\tau_{//}$) and out of plane ($\tau_\perp$) torques. An in-plane external magnetic field ($H_{ext}$) is applied at angle of φ with respect to the direction of RF current ($I_{rf}$). (b) The ST-FMR signal is measured at φ=45° using different driving frequencies from 6 to 16 GHz at T = 285 K. (c,d) The angular dependence taken using an excitation frequency of 10 GHz at room temperature shows the contribution of spin torques in all directions to extract the spin Hall coefficient using Eq. 3 in the main text. The symmetric (c) and anti-symmetric (d) rectifying voltages are shown as a function of φ. The solid blue, green and magenta lines are the fit curves corresponding to the torques acting in x, y and z directions. The red curve is the sum of all torques. Angular fitting of symmetric and antisymmetric coefficients using Eqs. 5 and 6 in the main text gives lower value of $\theta_{SH} = 0.35\pm0.1$ due to the smearing of topological nature of BSTS in direct contact with metallic ferromagnet Py.

## S3. Temperature dependent ST-FMR and NLV

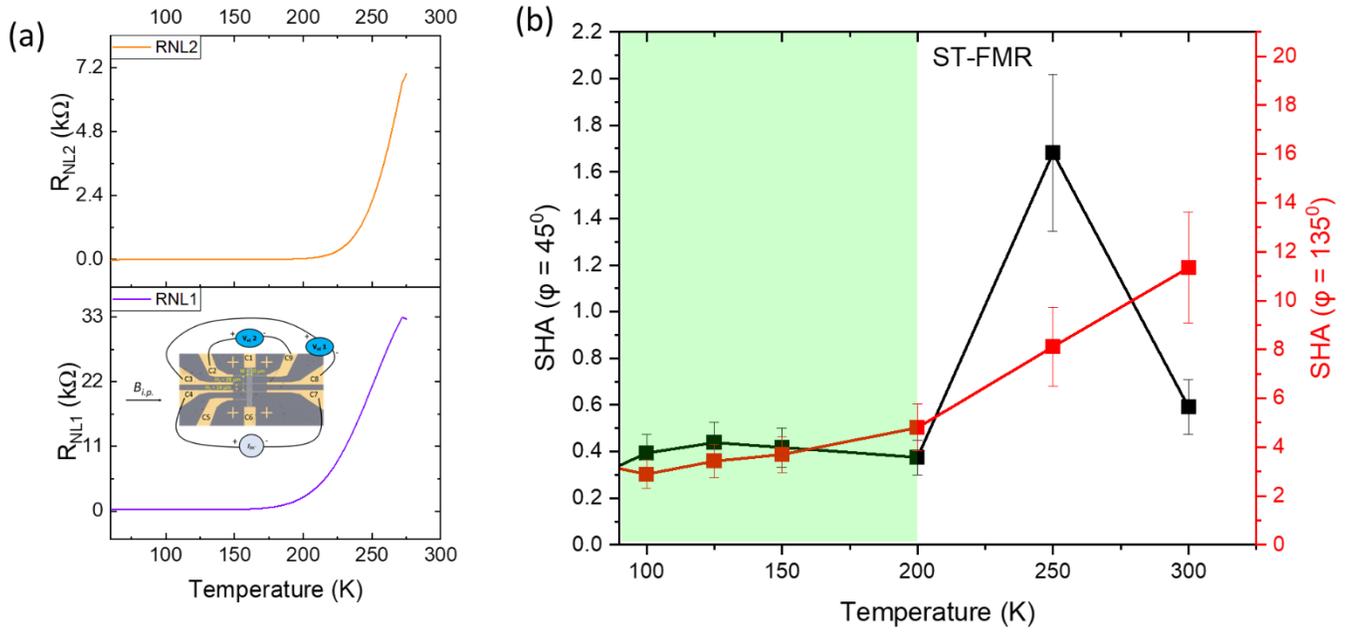

**Figure S3** (a) Temperature dependent zero field NLV measured at two different set of contacts. Non-local resistance RNL1 is between C2-C9 and RNL2 is between C3-C8. Zero field NLV voltage is reduced with a decrease in temperature due to the highly resistive nature of BSTS at low temperatures. Inset shows the measurement circuit for NLV. (b) Calculated ST-FMR spin Hall angles at 6GHz measured at different temperatures with magnetic field aligned at $45^0$ and $135^0$. A similar trend of decrease in spin Hall angle is observed.

## S4. Non-local voltage closer to current injection

The non-local voltages $V_{nl-1}$ and $V_{nl-2}$ are monitored at two different pairs of contacts as shown in the schematic of **Figure S4 (a)**. The data in the main text is from NL-2. The data for NL-1 is shown here for completeness. In **Figure S4 (b)**, we show the non-local resistance measured at NL-1 at different current bias while sweeping the in-plane magnetic field. We see the decrease in the non-local signal as current bias is increased and parabolic background from charge carriers bleeding into the nearby non-local voltage probes. The zero-field non-local resistance from both locations is shown in **Figure S4 (c)**. The drastic change in NL-1 as current increases above 5 µA showcases the contribution of charge current bleeding into the signal and suppressing the non-local voltage. It suggests that since the $\frac{H_1}{W} \sim 1$, the competition from charge carriers suppresses the spin transport. The data in **Figure S4 (b)** is fit using **Eq.2** of the main text and shown in **Figure S4 (d)**. For the signals at NL-1 shown here, we notice, $\theta_{SH} = 2 \pm 0.3$. It is comparable to the $\theta_{SH} = 2.83 \pm 0.64$ for NL-2 given in the main text, whereas the spin diffusion length is halved.

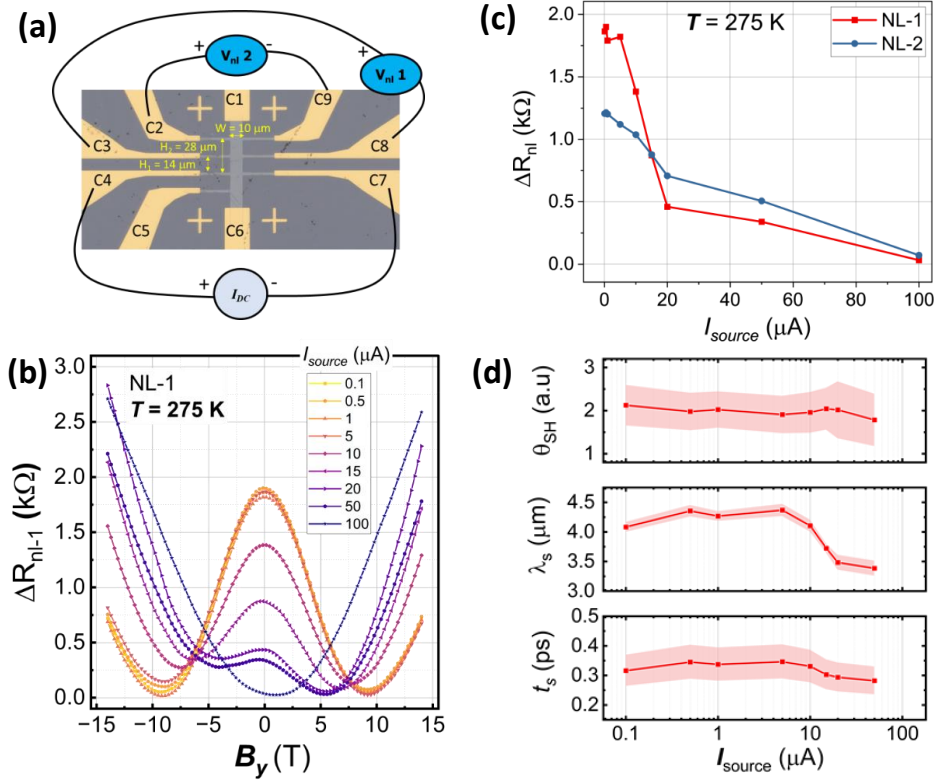

Figure S4. (a) Schematic showing all the connections for the non-local voltages at NL-1 ($H_1$ = 14 µm) and NL-2 ($H_2$ = 28 µm). (b) The non-local resistance measured at NL-1 at different bias. (c) The non-local

resistances at NL1- and NL-2 taken at $B_y = 0$ T show a decreasing trend with increased bias. The parameters from fitting the curves in (b) are shown in (d).

## S5. Fitting ST-FMR $V_{mix}$ signal to extract symmetric and anti-symmetric Lorentzian components

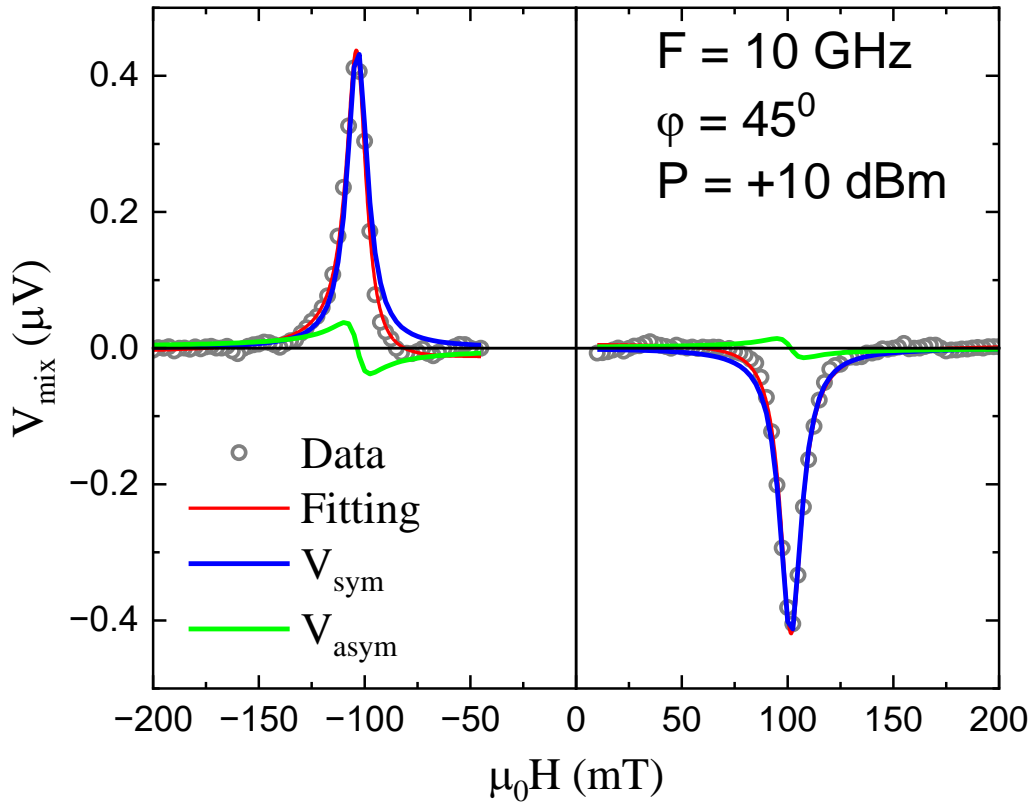

Figure S5. Exemplary ST-FMR signal at 10 GHz in BSTS/MgO/Py sample at room temperature. Data is fitted to the Eq. 3 given in the main text to extract symmetric and anti-symmetric Lorentzian components.

## S6. Kittel and LLG fits as a function of frequency

**Figure S6(a)** shows the excitation frequency as a function of resonance field for the ST-FMR devices with and without MgO interface. A witness sample of bare Py is also included for comparison. Fitting the data using the Kittel equation, $F = \frac{\gamma}{2\pi}\sqrt{H(H + \mu_0 M_{eff})}$, where $\gamma$ is the gyromagnetic ratio, yields us the effective magnetization parameter $\mu_0 M_{eff} \sim 7.53 \pm 0.5$ kOe for each sample. We also extract the Gilbert damping parameter ($\alpha$) and the inhomogeneous broadening parameter ($\Delta H_0$) using Landau-Lifshitz Gilbert model (LLG) by fitting the extracted ST-FMR linewidths ($\Delta H$) as a function of excitation frequency (**Figure S6(b)**), $\Delta H = \Delta H_0 + \frac{4\pi f \alpha}{\gamma}$. The FMR linewidth trend depicted in Fig. S6(b) shows that the sample InP/BSTS/Py does not exhibit drastic change in Gilbert damping as compared to InP/Py. However, the sample with MgO tunnel barrier (InP/BSTS/MgO/Py) clearly shows the change in FMR linewidth slope which is a signature of enhanced gilbert damping. This behavior is closely resembles the differences between spin Hall coefficient in the two samples.

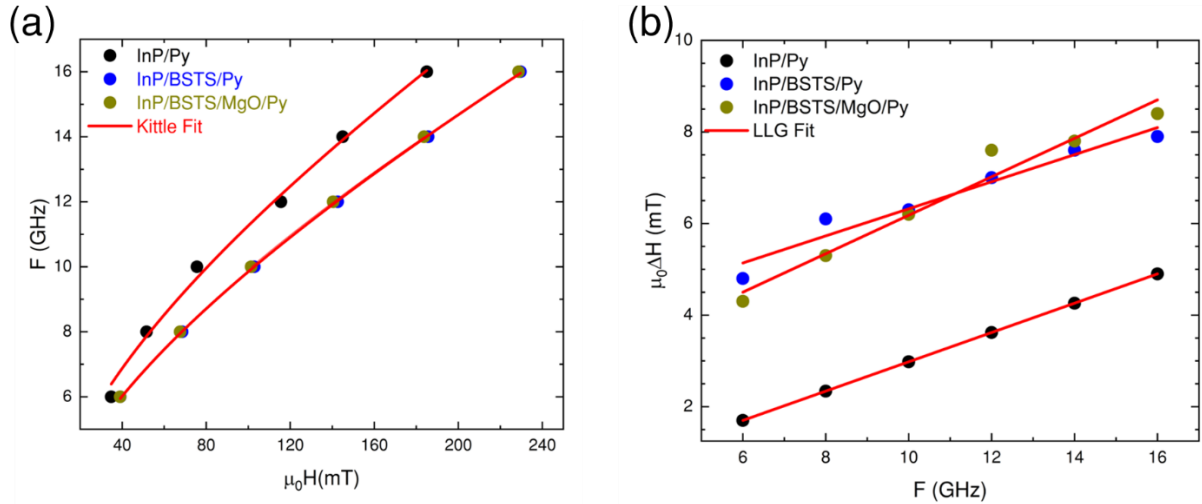

**Figure S6 (a)** Excitation frequency as a function of resonance field plotted (solid symbols) for three samples. Data is fitted (Red line) with Kittel equation to extract effective magnetization of the samples. (b) Extraction of Gilbert damping parameter using Landau-Lifshitz-Gilbert (LLG) fitting of ST-FMR linewidths for three samples.